\documentclass{iopart}
\usepackage{graphicx}  
\usepackage{latexsym}
\usepackage{amssymb}

\jl{6}        % submitted to journal: Classical and Quantum Gravity
\eqnobysec    % section equation numbering

\def\beq{\begin{equation}}
\def\eeq{\end{equation}}
\def\rmd{{\rm d}}

\def\leftcontractp#1{\mathop{\rlap{\hbox to 8pt{\hss$^{#1}$\hss}}%
  \hbox{\vrule height0.5pt width8pt \vrule width0.5pt  height6pt}}}
 
\def\rightcontractp#1{\mathop{%
  \hbox{\vrule width0.5pt height6pt \vrule height0.5pt width8pt}%
    \llap{\hbox to 8pt{\hss$^{#1}$\hss}}}}
 
\def\hook{\,\mathop{\breve{\,}}\nolimits\,}
\def\lefthook{\hook\kern-1.5pt}

\begin{document}

\title[Observer-dependent tidal indicators in the Kerr spacetime]
{Observer-dependent tidal indicators in the Kerr spacetime}

\author{
Donato Bini$^* {}^\S{}^\dag$ and
Andrea Geralico${}^\S{}^\ddag$
}

\address{${}^*$\
Istituto per le Applicazioni del Calcolo ``M. Picone,'' CNR, 
I--00185 Rome, Italy}

\address{
  ${}^\S$\
  ICRA,
  University of Rome ``La Sapienza,'' I--00185 Rome, Italy
}

\address{
  ${}^\dag$\
  INFN, Sezione di Firenze, I--00185
  Sesto Fiorentino (FI), Italy
}

\address{
  $^\ddag$
  Physics Department,
  University of Rome ``La Sapienza,'' I--00185 Rome, Italy}

\begin{abstract}
The observer-dependent tidal effects associated with the electric and magnetic parts of the Riemann tensor with respect to an arbitrary family of observers are discussed in a general spacetime in terms of certain \lq\lq tidal indicators.''
The features of such indicators are then explored by specializing our considerations to the family of stationary circularly 
rotating observers in the equatorial plane of the Kerr spacetime.
There exist a number of observer families which are special for several reasons and for each of them such indicators are evaluated.
The transformation laws of tidal indicators when passing from one observer to another are also discussed, clarifying the interplay among them.
Our analysis shows that no equatorial plane circularly rotating observer in the Kerr spacetime can ever measure a vanishing tidal electric indicator, whereas the family of Carter's observers measures zero tidal magnetic indicator.
\end{abstract}

\pacno{04.20.Cv}

\section{Introduction}

Tidal effects are commonly associated with spacetime curvature properties. 
In particular, when the spacetime is that of a black hole, such effects result in deforming and even tidally disrupting astrophysical objects, like ordinary stars but also compact objects like neutron stars. It is also widely believed that such tidal disruption events happen very frequently in the Universe, leading then to a possible detection of the associated emission of gravitational waves in the near future by ground-based detectors such as LIGO or VIRGO and their advanced versions \cite{abadieetal}.
Motivated by this interesting perspective, there has been in the recent literature a big effort to study tidal effects in many contexts, especially the merging of compact object binaries.
In fact, the presence of tidal interactions during the inspiral phase of neutron star-neutron star (NSNS) and black hole-neutron star (BHNS) binaries affects both the dynamics of the inspiral and the emitted waveform.
The techniques adopted range from Newtonian gravity to full general relativity, passing through Post-Newtonian approaches and numerical relativity simulations \cite{newt1,approxgr,fullgr1,fullgr2,fullgr3}.

NSNS and BHNS binary mergers can be accurately modeled only by numerical simulations in full general relativity, solving the coupled Einstein-hydrodynamics equations. 
However, there is a variety of analytical and semi-analytical approaches which allow to properly describe at least part of the coalescence process and to study the associated gravitational wave signals.
Flanagan and Hinderer \cite{fla-hin,hin1} adopted a post-Newtonian based description of the binary dynamics to study the influence of tidal effects in inspiralling BHNS systems. 
Within the Newtonian theory of tidal Love numbers \cite{love} they estimated the tidal responses of a neutron star to the external tidal solicitation of its companion, potentially measurable in gravitational wave signals from inspiralling binary neutron stars by Earth-based detectors.
The corresponding relativistic theory of Love numbers has been developed by Binnington and Poisson \cite{Bin-poi}.
An improvement of the analysis done in Ref. \cite{fla-hin}, whose validity was limited only to the early (lower-frequency) portion of the gravitational wave inspiral signal, was presented by Damour and Nagar \cite{damour1}, who used an extension of the so called ``effective one body approach'' to describe with sufficient accuracy the binary dynamics including tidal effects and covering the whole frequency range of the gravitational wave signal.

According to the standard theory of tides, the motion of an extended body in a given gravitational field is studied under the assumption that it causes only a small perturbation on the background, so that the emission of gravitational radiation are properly discussed in the framework of first order perturbation theory (see, e.g., Ref. \cite{mash75}).
Obviously the explicit expression of the tidal forces depends on the nature of the body under consideration, i.e., on the particular form of the energy-momentum tensor describing it.
For instance, one can assume a perfect fluid.
In general the energy momentum tensor can be determined by the complete set of its multipole moments with respect to a world line representing the motion of the body, according to Dixon's model \cite{dixon64,dixon69,dixon70,dixon73,dixon74}. 

Relativistic tidal problems have been extensively studied in the so-called tidal approximation, i.e., by assuming that the mass of the star is much smaller than the black hole mass and that the stellar radius is smaller than the orbital radius, thus neglecting backreaction effects.
In this approximation, the center of mass of the star is assumed to move around the black hole along a timelike geodesic path and the tidal field due to the black hole is computed from the Riemann tensor in terms of the geodesic deviation equation.
The internal stresses are usually accounted under the approximation that the classical laws of motion hold for the different points of the body relative to the center of mass, when the external tidal forces are neglected. 
Therefore, the body is usually described as a self-gravitating Newtonian fluid where the gravitational potential
associated with the tidal potential is linearly superposed to the internal forces.
Such a framework has been widely used to study the tidal disruption limit of ordinary stars and compact objects \cite{fishbone,hiscock,marck,carter,marck2,shibata,frolov,mino}.

If the dimensions of the body are assumed to be much smaller than the characteristic length scale associated with the background field, the motion of the extended body can be also described by the Mathisson-Papapetrou-Dixon equations \cite{math37,papa51,dixon64}. 
According to these, when the structure of the body is given by its spin only the deviation from geodesic motion  is due to the magnetic part of the Riemann tensor \cite{mashsingh,spindev_schw,spindev_kerr}.
Alternatively, the extended body can be represented by a collection of pointlike particles each following a timelike path.
The separation between a reference world line within the congruence (say that of a \lq\lq fiducial observer'') and a generic one is represented by a connecting vector, whose components with respect to an orthonormal frame adapted to the observer satisfy a second order deviation equation \cite{strains1}.
In the case of a geodesic congruence one can set up a parallel transported frame so that the geodesic deviation equation takes a form reminiscent of a Newtonian equation \cite{pirani56}, where tidal forces are due to a Newtonian quadrupole field.
The electric part of the Riemann tensor represents the only direct curvature contribution to the geodesic deviation equation (as well as the deviation equation for general non-geodesic motion), introducing shearing forces.
Such an approach has been used to study tidal dynamics of relativistic flows in the Kerr spacetime for different choices of the reference world line \cite{mcclune,chicmash1,chicmash2,strains2}.

The aim of the present paper is to study the role of the observer in relativistic tidal problems and to provide all necessary tools to relate the measurement of tidal effects by different families of observers.  
In fact, if tidal forces are due to curvature, the latter is experienced by observers, i.e., test particles or extended bodies (hereafter simply \lq\lq observers"), through the electric and magnetic parts of the Riemann tensor.
The Riemann tensor is indeed the true  $4$-dimensional invariant quantity, whereas the various parts mentioned above depend on the choice of the observers who perform the measurement. 
We will start by examining in detail all these definitions and notions, including the corresponding transformation laws arising when changing the observer in full generality.
We will consider then two \lq\lq tidal indicators" defined as the trace of the square of the electric and magnetic and mixed parts of the Riemann tensor, respectively.
They are both curvature and observer dependent and we will explore how they change when the observer also changes.  

To be more explicit and to relate our work to other existing studies, we discuss the case of a Kerr spacetime and the family of circularly rotating (accelerated) observers in the equatorial plane.
There exists a whole collection of geometrically special circular orbits \cite{idcf1,idcf2,bjm,bjdf} in the equatorial plane of the Kerr spacetime which arise in the discussion of some geometrical and physical features (i.e., transport laws for vectors, geodesics and accelerated orbits, clock effects, circular holonomy, etc.).
They have been classified in Ref. \cite{bjelba} with respect to their geometrical meaning as follows:
extrinsically special orbits, in the sense that their specialness comes from the
background geometry;
intrinsically special orbits, in the sense that their specialness comes from the
intrinsic Frenet-Serret properties of the curve itself;
relatively special orbits, in the sense that their specialness is with respect to other
curves.
We will investigate the properties of tidal indicators as associated with each special family of observers mentioned above, so providing a complete analysis of tidal forces.
Previous works limited their considerations only to geodesic observers, thus giving only a partial insight into the tidal problem. 
Relative-observer tidal indicators should play instead a more central role.
Furthermore, special observers can be identified as suitable to study relativistic tidal dynamics.

Hereafter latin indices run from $1$ to $3$ whereas greek indices run from $0$ to $3$ and geometrical units are assumed. The metric signature is chosen 
as $+2$.

\section{Electric and magnetic parts of the Weyl tensor}

The electric and magnetic parts of the Weyl tensor $C_{\alpha\beta\gamma\delta}$ with respect to a generic timelike congruence with unit tangent vector $u$ are defined as \cite{book}
\beq
\label{weylsplitu}
E(u)_{\alpha\beta}=C_{\alpha\mu\beta\nu}u^\mu u^\nu\,,\qquad 
H(u)_{\alpha\beta}=-C^*_{\alpha\mu\beta\nu}u^\mu u^\nu\,.
\eeq 
These spatial fields are both symmetric and tracefree.
We recall that in a vacuum spacetime, which is the case we will consider below for applications, the Weyl and Riemann tensors coincide.

Let $U$ be another unit timelike vector field representing a different family of observers related to $u$ by a boost, i.e.,
\beq
\label{Usplitcap7}
U=\gamma(U,u)[u+\nu(U,u)]\,, \qquad
\gamma(U,u)=1/\sqrt{1-||\nu(U,u)||^2}\,.
\eeq
Both observers can be used to split the Weyl tensor in terms of associated electric and magnetic parts, so that a decomposition analogous to Eq. (\ref{weylsplitu}) holds also for $U$, i.e.,
\beq
\label{weylsplit2}
E(U)_{\alpha\beta}= C_{\alpha\mu\beta\nu}U^\mu U^\nu\,, \qquad
H(U)_{\alpha\beta}=-C^*_{\alpha\mu\beta\nu}U^\mu U^\nu\,.
\eeq 
Using (\ref{Usplitcap7}) we then have 
\beq
E(U)_{\alpha\beta}=\gamma(U,u)^2 C_{\alpha\mu\beta\nu}[u^\mu+\nu(U,u)^\mu] [u^\nu+\nu(U,u)^\nu]\,,
\eeq
that is, introducing the notation $\gamma(U,u)=\gamma$ and $\nu(U,u)=\nu$, 
\begin{eqnarray}
\fl\quad
E(U)_{\alpha\beta}=\gamma^2 [E(u)_{\alpha\beta}+C_{\alpha\mu\beta\nu}u^\mu \nu^\nu 
+C_{\alpha\mu\beta\nu} \nu ^\mu u^\nu +C_{\alpha\mu\beta\nu}\nu^\mu \nu ^\nu]\,.
\end{eqnarray}
These relations can be better represented in the frame language. 
Let $\{e_\alpha\}$ denote an adapted frame to $u$, namely $u=e_0$ and the spatial triad $\{e_a\}$ spans the local rest space of $u$. 
The frame components of $E(U)$ are given by (see also Ref. \cite{maar} and references therein for a recent review on this and related topics)
\begin{eqnarray}
\label{EUframecomp}
\fl\quad
E(U)_{ab}&=&\gamma^2 [E(u)_{ab}+C_{a0 bc} \nu^c +C_{ac b0} \nu ^c  +C_{ac bd}\nu^c \nu^d]\nonumber \\
\fl\quad
&=& \gamma^2 [E(u)_{ab}+H(u)_{af}\eta(u)^f{}_{bc} \nu^c +H(u)_{bf}\eta(u)^f{}_{ac} \nu ^c\nonumber \\
\fl\quad
&&  -E(u)_{fg}\eta(u)^f{}_{ac}\eta(u)^g{}_{bd}\nu^c \nu^d]\,,\nonumber \\
\fl\quad
E(U)_{00}&=&\gamma^2 [C_{0b0c}\nu^b \nu^c]=\gamma^2 [E(u)_{bc}\nu^b \nu^c]\,,\nonumber \\
\fl\quad
E(U)_{0b}&=&\gamma^2 [C_{0c b0} \nu ^c  +C_{0c bd}\nu^c \nu^d]=\gamma^2 [-E(u)_{cb}  \nu ^c -H(u)_{cf}\eta(u)^f{}_{bd}\nu^c \nu^d]\,,
\end{eqnarray}
where the spatial unit-volume $3$-form has been introduced, i.e.,
\beq
\eta(u)_{\alpha\beta\gamma}=u^\mu \eta_{\mu\alpha\beta\gamma}\,,\qquad \eta(u)_{abc}=\eta_{0abc}.
\eeq
Relations (\ref{EUframecomp}) are still rather involved to be of practical use.
A first progress is obtained by introducing the magnitude and the spatial dual (i.e., a ${}^{*(u)}$-operation defined with $\eta(u)_{abc}$) of the velocity vector, that is
\beq
\nu^c= \nu \,\hat\nu^c\,,\qquad \nu \, \hat V^{ab}=\nu \, \eta(u)^{ab}{}_c \hat\nu^c\equiv \nu [{}^{*(u)}\hat \nu]^{ab}\,,
\eeq
with $\hat V^{ab}=-\hat V^{ba}$.
We find
\begin{eqnarray}
\label{EUframecomp2}
\fl\quad
E(U)_{ab}
&=& \gamma^2 \left[E(u)_{ab}+\nu [H(u)\hat V]_{ab} -\nu [\hat VH(u)]_{ab}+\nu^2 [\hat VE(u)\hat V]_{ab}\right]\,, \nonumber\\
\fl\quad
E(U)_{00}&=&\gamma^2\nu^2  [E(u)_{bc}\hat \nu^b \hat \nu^c]\,,\nonumber \\
\fl\quad
E(U)_{0b}
&=& -\gamma^2\nu \hat \nu ^c \left[E(u)_{cb}  + \nu [H(u)\hat V]_{cb}\right]\,.
\end{eqnarray}

Analogous expressions hold for $H(U)$ (simply obtained by the replacements $E(U) \to H(U)$ and 
$H(U)\to -E(U)$, because of the duality property ${}^*C^*=-C$ of the Weyl tensor).

\section{Circular equatorial orbits in the Kerr spacetime}
 
In standard Boyer-Lindquist coordinates the Kerr metric writes as
\beq\fl\quad
\rmd s^2 = -\left(1-\frac{2Mr}{\Sigma}\right)\rmd t^2 -\frac{4aMr}{\Sigma}\sin^2\theta\rmd t\rmd\phi+ \frac{\Sigma}{\Delta}\rmd r^2 +\Sigma\rmd \theta^2+\frac{\Lambda}{\Sigma}\sin^2 \theta \rmd \phi^2\ ,
\eeq
where $\Delta=r^2-2Mr+a^2$, $\Sigma=r^2+a^2\cos^2\theta$ and $\Lambda = (r^2+a^2)^2-\Delta a^2\sin^2 \theta$. Here $a$ and $M$ are the specific angular momentum and total mass characterizing the spacetime. The event horizons are located at $r_\pm=M\pm\sqrt{M^2-a^2}$. 

Introduce the zero angular momentum observer (ZAMO) family of fiducial observers with 4-velocity $n$ orthogonal to the time coordinate hypersurfaces
\beq
\label{n}
n=N^{-1}(\partial_t-N^{\phi}\partial_\phi)\ ,
\eeq
where $N=(-g^{tt})^{-1/2}=\left[\Delta\Sigma/\Lambda\right]^{1/2}$ and $N^{\phi}=g_{t\phi}/g_{\phi\phi}=-2aMr/\Lambda$ are the lapse function and only nonvanishing component of the shift vector field respectively. 
The ZAMOs are accelerated and locally nonrotating in the sense that their vorticity vector vanishes; they have also a nonzero expansion tensor.
A suitable orthonormal frame adapted to the ZAMOs is given by
\beq
\label{zamoframe}
e_{\hat t}=n\ , \quad
e_{\hat r}=\frac1{\sqrt{g_{rr}}}\partial_r\ , \quad
e_{\hat \theta}=\frac1{\sqrt{g_{\theta \theta }}}\partial_\theta\ , \quad
e_{\hat \phi}=\frac1{\sqrt{g_{\phi \phi }}}\partial_\phi\ .
\eeq
When referring to them we also use the notation $e_{0}=e_{\hat t}$, $e_{\hat r}=e_{1}$, $e_{\hat \theta}=e_2$ and $e_{\hat \phi}=e_3$.

Consider a family of uniformly rotating timelike circular orbits at a given fixed radius on the equatorial plane with  4-velocity vector $U$. It can be parametrized equivalently either by the constant angular velocity $\zeta$ with respect to infinity or by the constant relative velocity $\nu$ with respect to the ZAMOs (defining the usual Lorentz factor $\gamma=(1-\nu^2)^{-1/2} $) or by the constant ZAMO rapidity $\alpha$ as follows 
\beq
\label{orbita}
U=\Gamma [\partial_t +\zeta \partial_\phi ]
 =\gamma [n +\nu e_{\hat \phi}]
 =\cosh\alpha \, n + \sinh\alpha \, e_{\hat\phi}
\ ,
\eeq
where $\Gamma$ is a normalization factor such that $U_\alpha U^\alpha =-1$ and hence
\beq
\Gamma =\left[ N^2-g_{\phi\phi}(\zeta+N^{\phi})^2 \right]^{-1/2}
       =\gamma/N
\eeq
with
\beq
\zeta=-N^{\phi}+(N/\sqrt{g_{\phi\phi}})\, \nu\ ,\quad
\nu=\sqrt{g_{\phi\phi}} (\zeta+N^{\phi})/N =\tanh \alpha\ .
\eeq

The Boyer-Lindquist coordinates in which the metric is commonly written are adapted to the Killing symmetries of the spacetime itself and automatically select the family of static or \lq\lq threading'' observers, i.e. those at rest with respect to the coordinates, following the time coordinate lines. 
The angular velocity of the threading observers is $\zeta_{\rm(thd)}=0$, whereas their relative velocity with respect to ZAMOs is
\beq
\nu_{\rm(thd)}=-\frac{2aM}{r\sqrt{\Delta}}\ .
\eeq
ZAMOs are instead characterized by
\beq
\zeta_{\rm (zamo)}=\frac{2aM}{r^3+a^2r+2a^2M}\,, \qquad
\nu_{\rm(zamo)}=0\ .
\eeq

A family of spatially circular orbits with particular properties was found by Carter \cite{carter68}. 
Their tangent vector is given by
\beq
\label{carter}
u_{\rm (car)}= \frac{r^2+a^2}{\sqrt{\Delta \, \Sigma}} \left[\partial_t + \frac{a}{r^2+a^2}\partial_\phi \right]\,, 
\eeq
so that
\beq
\zeta_{\rm (car)}=\frac{a}{r^2+a^2}\,, \qquad
\nu_{\rm (car)}=\frac{a\sqrt{\Delta}}{r^2+a^2}\,.
\eeq
The main property of these trajectories is to be the unique timelike world lines belonging to the intersection of the Killing two-plane ($t,\,\phi$) with the two-plane spanned by the two independent  principal null directions of the Kerr spacetime. 
Another special property of these observers is that the components of a geodesic
4-velocity in a frame adapted to the Carter observers separate in their dependence on
the $r$ and $\theta$ coordinates, allowing the exact integration of the geodesic equations of
motion.
Furthermore, the electric and magnetic parts of the Weyl tensor as measured by them
are aligned.

Natural extrinsically special timelike circular orbits on the equatorial plane are
the co-rotating $(+)$ and counter-rotating $(-)$ geodesics (with respect to the hole rotation,
which is clockwise assuming $a > 0$) characterized by the following angular and linear velocities
\beq\fl\quad
\zeta_{({\rm geo})\, \pm}\equiv\zeta_{\pm}
=\left[a\pm (M/r^3)^{-1/2}\right]^{-1}\ , \
\nu_{({\rm geo})\, \pm}\equiv \nu_\pm 
=\frac{a^2\mp2a\sqrt{Mr}+r^2}{\sqrt{\Delta}(a\pm r\sqrt{r/M})}\ ,
\eeq 
respectively.
The corresponding timelike conditions $|\nu_\pm|<1$ identify the allowed regions $r>r_{{(\rm geo)}\pm}$ for the radial coordinate where co/counter-rotating geodesics exist, where the null circular orbits occur at 
\beq
r_{{(\rm geo)}\pm}
=2M\left\{1+\cos\left[\frac23\arccos\left(\pm\frac{a}{M}\right)\right]\right\}\ .
\eeq

Closely related to these are the ``geodesic meeting point observers'' defined by their alternating successive intersection points \cite{idcf1,idcf2}, with angular and linear velocities given by
\beq\fl
\zeta_{\rm (gmp)}=\frac{
\zeta_{+}+\zeta_{-}}{2}=-\frac{aM}{r^3-a^2M}\,,\qquad
\nu_{\rm (gmp)}=\frac{
\nu_{+}+\nu_{-}}{2}=-\frac{aM(3r^2+a^2)}{\sqrt{\Delta}(r^3-a^2M)}\ ,
\eeq
respectively.
These observers enter the discussion of the general relativistic definition of inertial
forces, clock-effects, and holonomy invariance.

The main geometrical and kinematical features of accelerated orbits with special geometrical properties are better discussed using the Frenet-Serret  formalism \cite{iyer-vish,bmj}, which reflects only the geometrical properties of spacetime and of the world line, being defined in an invariant manner without reference to any particular coordinate system or observer.
For accelerated circular orbits in the equatorial plane the relevant Frenet-Serret intrinsic quantities are the magnitude of the acceleration $\kappa$ and the first torsion $\tau_1$
\begin{eqnarray}
\kappa&=&k_0\gamma^2 (\nu-\nu_+)(\nu-\nu_-)\ ,\qquad \tau_1= -\frac{1}{2\gamma^2}\frac{d\kappa}{d {\nu}}\,,
\end{eqnarray} 
with 
\beq
k_0=-\frac{\sqrt{\Delta}(r^3-a^2M)}{r^2(r^3+a^2r+2a^2M)}\,,
\eeq
whereas the second torsion vanishes identically.
Therefore, one can study how the angular velocity (or linear velocity) affects $\kappa$ and $\tau_1$, since they are both functions of $\nu$.
Thus apart from the case of geodesics for which $\kappa_{\rm(geo)\pm} = 0$ and the Frenet-Serret approach degenerates, the obvious intrinsically
special orbits on the equatorial plane of the Kerr spacetime correspond to $\tau_1 = 0$ or equivalently to $\rmd\kappa/\rmd\nu=0$ and are therefore called either ``critically accelerated'' or ``extremely accelerated'' orbits \cite{idcf1,idcf2,bjm,bjdf,bjelba,iyer-vish,fdfacc,semerak}, since they extremize the curvature $\kappa$ among the family of circular orbits.
In the equatorial plane of the Kerr spacetime there are two critically accelerated linear velocities $\nu_{{\rm (crit)}\pm}$, related to the geodesic velocities by
\begin{eqnarray}\fl\quad
\nu_{{\rm (crit)}\pm}&=&\frac{\gamma_- \nu_- \mp \gamma_+ \nu_+}{\gamma_- \mp \gamma_+} \nonumber \\
\fl\quad
&=& -\frac1{2Ma(3r^2+a^2)\sqrt{\Delta}}\Big[-2a^2M(a^2-3Mr)+r^2(r^2+a^2)(r-3M)\nonumber\\
\fl\quad
&&\pm(r^3+a^2r+2a^2M)\sqrt{r}\sqrt{r(r-3M)^2-4a^2M}\Big]\ .
\end{eqnarray}
In the literature the name extremely accelerated observer is reserved for the 
$U_{\rm(crit)-}$ (i.e. $U_{\rm(ext)}=U_{\rm(crit)-}$) ,
because this 4-velocity is timelike far enough from the hole where the counter- and co-rotating geodesics are both timelike, while the other one is spacelike there.
In fact, $|\nu_{{\rm (crit)}-}|<1$ in both regions $r_+<r<r_{{(\rm geo)}+}$ and $r>r_{{(\rm geo)}-}$ where $|\nu_{{\rm (crit)}+}|>1$, while $|\nu_{{\rm (crit)}+}|<1$ holds for the complementary region $r_{{(\rm geo)}+}<r<r_{{(\rm geo)}-}$. 
The extremely accelerated observers play a central role in the discussion of general relativistic gravitomagnetic clock effects.

\begin{figure} 
\begin{center}
\includegraphics[scale=.35]{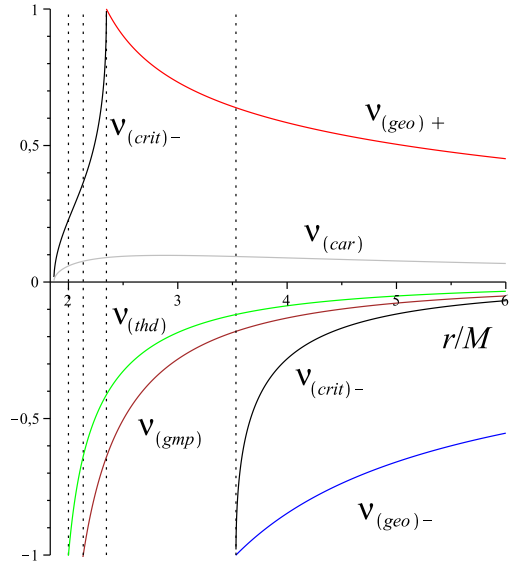}
\end{center}
\caption{
The behavior of the linear velocity as a function of the radial coordinate corresponding to the various observers introduced in section 2 is shown for the choice $a/M=0.5$. Dashed  vertical lines correspond to observer horizons.
} 
\label{fig:1}
\end{figure}

Fig. \ref{fig:1} shows the behavior of the linear velocity for the various observers introduced above as a function of the radial coordinate.

% figure 1

\section{Tidal indicators}

Tidal forces are commonly associated with the Riemann tensor and more specifically with its  electric and magnetic parts with respect to a generic timelike congruence. Let us denote by $u$
the corresponding unit tangent vector. 
For instance, when one studies the geodesic deviation among a set of test particles the role of an interaction force field is played by $E(u)$; if the particles are endowed with a spinning structure, the deviation from geodesic behavior is instead due to $H(u)$. So both of them are relevant when considering tidal problems.

The simplest way to built up scalar quantities through the electric and magnetic parts of the Riemann tensor which are representative of them and serve as \lq\lq tidal indicators" in the study of tidal effects is to take the trace of their square.
One can then consider the following electric and magnetic tidal indicators
\beq
\label{tidalind}
{\mathcal T}_E(u)= {\rm Tr}[E(u)^2]\,,\qquad 
{\mathcal T}_H(u)= {\rm Tr}[H(u)^2]\,.
\eeq
They are both related to the curvature tensor and to the particle/observer undergoing tidal deformations.
Other more involved tidal indicators have been discussed in Ref. \cite{mino}, which are but of interest only when using Fermi coordinate tidal potential. 

Let $u=n$ be the unit tangent vector to the ZAMO family of observer given by Eq. (\ref{n}) with adapted frame (\ref{zamoframe}).
The relevant nonvanishing frame components of the electric and magnetic parts of the Riemann tensor are given by 
\begin{eqnarray}
E(n)_{11}&=&\frac{M}{r^4}\frac{2a^2Mr-3a^4-5r^2a^2-2r^4}{2a^2M+r^3+a^2r}\,, \nonumber\\
E(n)_{22}&=&-\frac{M}{r^4}\frac{4a^2Mr-3a^4-4r^2a^2-r^4}{2a^2M+r^3+a^2r}\,, \nonumber\\
H(n)_{12}&=&-\frac{3aM}{r^4}\frac{(r^2+a^2)\sqrt{\Delta}}{2a^2M+r^3+a^2r}\,,
\end{eqnarray}
with
\beq
E(n)_{11}+E(n)_{22}=-E(n)_{33}=-\frac{M}{r^3}\,;
\eeq
the tidal indicators (\ref{tidalind}) then  turn out to be given by
\begin{eqnarray}
\label{tidalindzamo}
{\mathcal T}_E(n)&=&\frac{6M^2}{r^6}+{\mathcal T}_H(n)\,,\nonumber\\
{\mathcal T}_H(n)&=&\frac{18a^2M^2}{r^8}\frac{(r^2+a^2)^2\Delta}{(2a^2M+r^3+a^2r)^2}\,.
\end{eqnarray}

Let now $U$ be tangent to a uniformly rotating timelike circular orbit on the equatorial plane of the Kerr spacetime, as given by Eq. (\ref{orbita}).
From the general transformation laws (\ref{EUframecomp2}) derived in section 2 we find
\begin{eqnarray}\fl\quad
\label{TrE2kerr}
{\rm Tr}[E(U)^2]&=&[E(U)_{00}]^2-2 E(U)_{0}{}^bE(U)_{0b}+E(U)^{ab}E(U)_{ab}\nonumber\\
\fl\quad
&=&2\gamma^4\left\{
([E(n)_{11}]^2+[E(n)_{22}]^2+E(n)_{11}E(n)_{22})(\nu^4+1)\right.\nonumber\\
\fl\quad
&&\left.-2H(n)_{12}(E(n)_{11}-E(n)_{22})\nu(\nu^2+1)\right.\nonumber\\
\fl\quad
&&\left.-\nu^2([E(n)_{11}]^2+[E(n)_{22}]^2+4E(n)_{11}E(n)_{22}-4[H(n)_{12}]^2)
\right\}
\,, 
\end{eqnarray}
and analogously for ${\rm Tr}[H(U)^2]$.
Eq. (\ref{TrE2kerr}) can also be written in a more compact form as 
\begin{eqnarray}\fl\quad
\label{TrE2kerr2}
{\rm Tr}[E(U)^2]&=&\gamma^4\left[
{\rm Tr}[E(n)^2](\nu^4+1)-4\nu(\nu^2+1){\rm Tr}[H(n)E(n)\hat V]\right.\nonumber\\
\fl\quad
&&\left.+\nu^2(-{\rm Tr}[E(n)^2]+3{\rm Tr}[({\hat V}E(n))^2]+4{\rm Tr}[H(n)^2])
\right]
\,,
\end{eqnarray}
where the antisymmetric matrix $\hat V$ has the only nonvanishing components $\hat V_{12}=1=-\hat V_{21}$.  

A convenient way to express the tidal indicators (\ref{tidalind}) showing their relative-observer properties is then the following
\begin{eqnarray}
\label{tidalind2}
{\mathcal T}_E(U)&=&\frac{6M^2}{r^6}\gamma_{\rm (car)}^4\gamma^4(\nu_{\rm (car)}^2-\nu_{\rm (car)}+1)(\nu_{\rm (car)}^2+\nu_{\rm (car)}+1)\nonumber\\
&&\times\left(\nu^2-\frac{2\nu}{\nu_{\rm(rel)-}}+1\right)\left(\nu^2+\frac{2\nu}{\nu_{\rm(rel)+}}+1\right)\,\nonumber\\
&=& {\mathcal T}_E(n)\gamma^4\left(\nu^2+\frac{2\nu}{\nu_{\rm(rel)-}}+1\right)\left(\nu^2-\frac{2\nu}{\nu_{\rm(rel)+}}+1\right)\,,\nonumber\\
{\mathcal T}_H(U)&=&\frac{18M^2}{r^6}\gamma_{\rm (car)}^4\nu_{\rm (car)}^2\gamma^4(\nu-\nu_{\rm (car)})^2(\nu-\bar\nu_{\rm (car)})^2\,\nonumber\\
&=& {\mathcal T}_H(n) \gamma^4(\nu-\nu_{\rm (car)})^2(\nu-\bar\nu_{\rm (car)})^2\,,
\end{eqnarray}
where 
\beq
\label{nureldef}
\nu_{\rm(rel)\pm}={2}\frac{\nu_{\rm (car)}^2\mp\nu_{\rm (car)}+1}{\nu_{\rm (car)}^2\mp4\nu_{\rm (car)}+1}\
\eeq
and we have used the notation $\bar\nu_{\rm (car)}=1/\nu_{\rm (car)}$ (see Appendix A for an equivalent parametrization in terms of the rapidity.)
As functions of $\nu_{\rm (car)}$, both $\nu_{\rm(rel)\pm}$ are such that $|\nu_{\rm(rel)\pm}|>1$. 
Note that the following relation holds
\beq
\label{relTEH}
{\mathcal T}_E(U)-{\mathcal T}_H(U)=\frac{6M^2}{r^6}\,.
\eeq
This invariance property has a simple explanation, since it is proportional to the Kretschmann invariant of the spacetime (evaluated on the equatorial plane)
\beq
K=C_{\alpha\beta\gamma\delta}C^{\alpha\beta\gamma\delta}\big|_{\theta=\pi/2}=\frac{48 M^2}{r^6}\,.
\eeq

\begin{figure}
\begin{center}
$\begin{array}{cc}
\includegraphics[scale=.28]{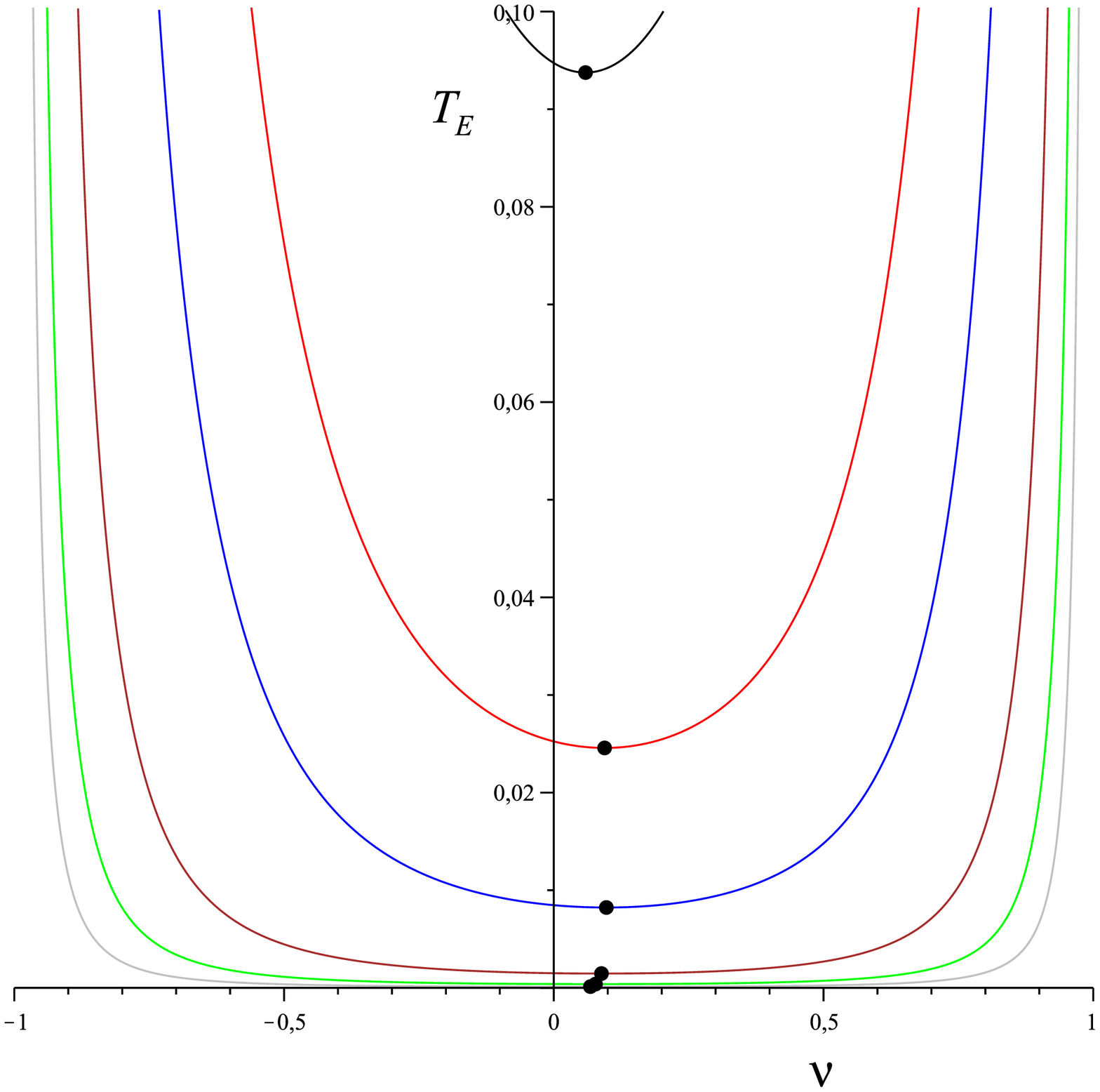}& 
\includegraphics[scale=.28]{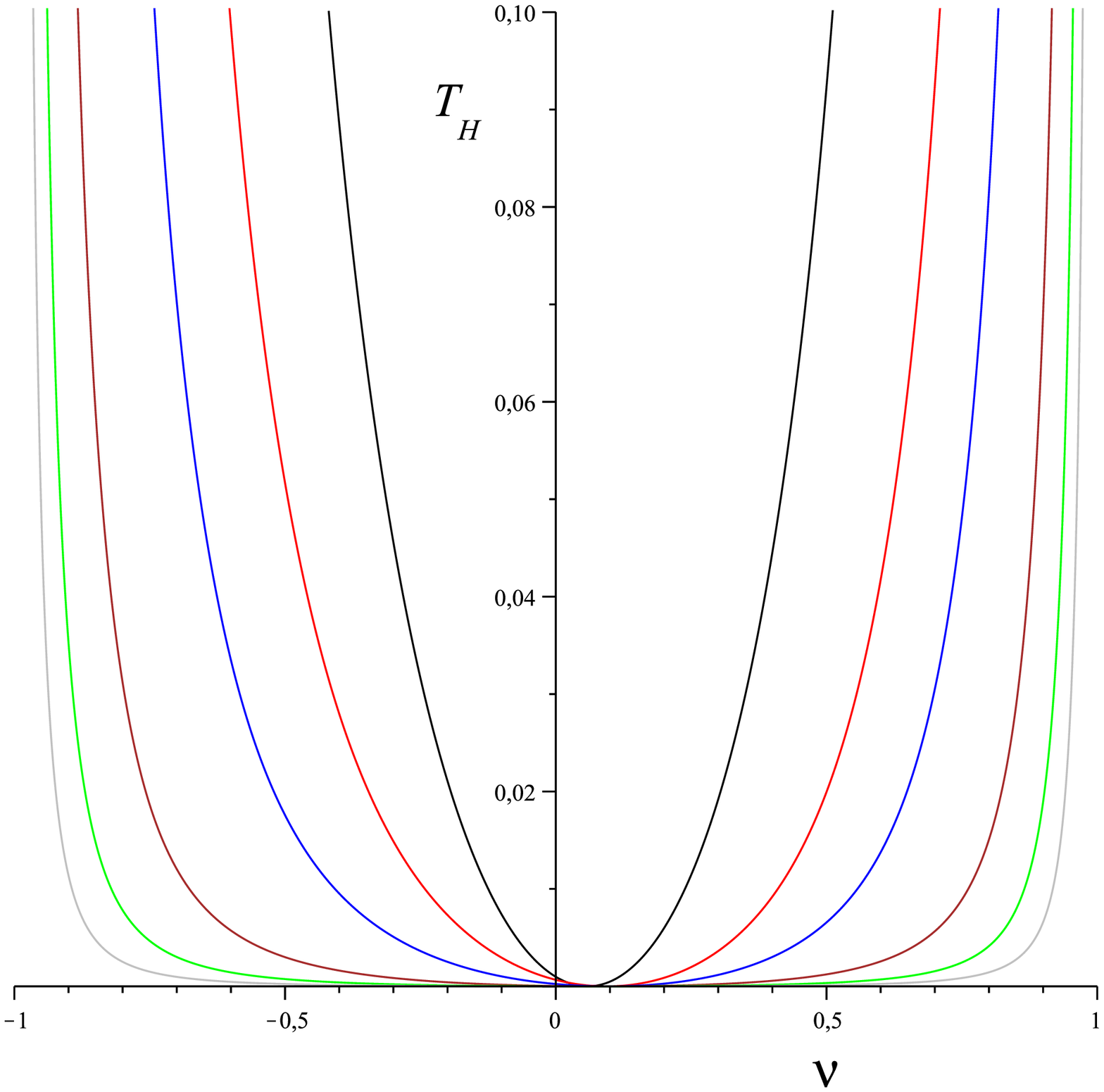}\\[0.4cm] 
\mbox{(a)}&\mbox{(b)}\\
\end{array}$
\end{center}
\caption {The behaviors of the tidal indicators ${\mathcal T}_E(U)$ and ${\mathcal T}_H(U)$ as functions of $\nu$ are shown in panels (a) and (b) respectively for the choice $a/M=0.5$ and different values of the radial coordinate $r/M=[2, 2.5, 3, 4, 5, 6]$. 
Dots in (a) show the minimum of each curve.
For increasing values of $r$ the curves open shrinking to the horizontal axis. 
Units on the vertical axis are chosen so that $M=1$.
} 
\label{fig:2}
\end{figure}

Eqs. (\ref{tidalind2}) and (\ref{relTEH}) imply that ${\mathcal T}_H(U)$ vanishes for $\nu=\nu_{\rm (car)}$, and correspondingly ${\mathcal T}_E(U)$ takes its minimum value. In fact, 
\beq\fl
\frac{\rmd {\mathcal T}_E(U)}{\rmd\nu}=\frac{36M^2}{r^6}\gamma_{\rm (car)}^4\nu_{\rm (car)}\gamma^6(\nu-\nu_{\rm (car)})(\nu-\bar\nu_{\rm (car)})[(\nu^2+1)(\nu_{\rm (car)}^2+1)-4\nu\nu_{\rm (car)}]\,.
\eeq
The behaviors of the tidal indicators as functions of $\nu$ are shown in Fig. \ref{fig:2}.

Interestingly the right hand side of Eq. (\ref{relTEH}) does not depend on the black hole rotation parameter $a$, implying that the difference between the two tidal indicators remains the same also in the Schwarzschild limit ($a=0$), where 
\beq\fl\qquad
{\mathcal T}_E^{\rm(schw)}(U)=\frac{6M^2}{r^6}\gamma^4(1+\nu^2+\nu^4)\,,\qquad
{\mathcal T}_H^{\rm(schw)}(U)=\frac{18M^2}{r^6}\gamma^4\nu^2\,.
\eeq

Fundamental in order to understand the dependence of ${\mathcal T}_E(U)$ and ${\mathcal T}_H(U)$ on the choice of the observer within the above specified class of observers is therefore the $\nu$-dependent factor which multiplies the corresponding ZAMO quantities ${\mathcal T}_E(n)$ and ${\mathcal T}_H(n)$ in Eqs. (\ref{tidalind2}).
It is evident that
${\mathcal T}_H(U)$ can be made vanishing (by selecting $\nu=\nu_{\rm (car)}$, a fact that gives Carter's observers a new characteristic  property), whereas ${\mathcal T}_E(U)$ is always different from zero for every choice of $\nu\in (-1,1)$, i.e., whoever is the observer measuring it.
This seems to be due to the very special nature of the Kerr solution, whose quadrupole moment is associated entirely with the angular momentum \cite{ht68}. 
We expect that the effect of a nonzero mass quadrupole moment  would significantly change the present results.
This further investigation will be considered in a future work.

Obviously our choice of ZAMOs as fiducial observers is just a possible choice and does not give a particular form to Eqs. (\ref{tidalind2}). In fact, the latter equations when referred to another observer family, say $u_1$, assume exactly the same form with $n$ replaced by $u_1$ and with the corresponding velocities transformed properly. For instance,
\begin{eqnarray}
\fl\quad
{\mathcal T}_H(U)&=& {\mathcal T}_H(u_1) \gamma(U,u_1)^4[\nu(U,u_1)-\nu(u_{\rm (car)},u_1)]^2[\nu(U,u_1)-\bar\nu(u_{\rm (car)},u_1)]^2\,,
\end{eqnarray}
where
\begin{eqnarray}
\nu(U,u_1)&=&\frac{\nu(U,n)-\nu(u_1,n)}{1-\nu(U,n)\nu(u_1,n)}\equiv \frac{\nu-\nu_1}{1-\nu \nu_1}\,,\nonumber \\
\nu(u_{\rm (car)},u_1)&=&\frac{\nu(u_{\rm (car)},n)-\nu(u_1,n)}{1-\nu(u_{\rm (car)},n)\nu(u_1,n)}\equiv 
\frac{\nu_{\rm (car)}-\nu_1}{1-\nu_{\rm (car)}\nu_1}\,.
\end{eqnarray}

\begin{figure} 
\typeout{*** EPS figure 3}
\begin{center}
$\begin{array}{cc}
\includegraphics[scale=0.28]{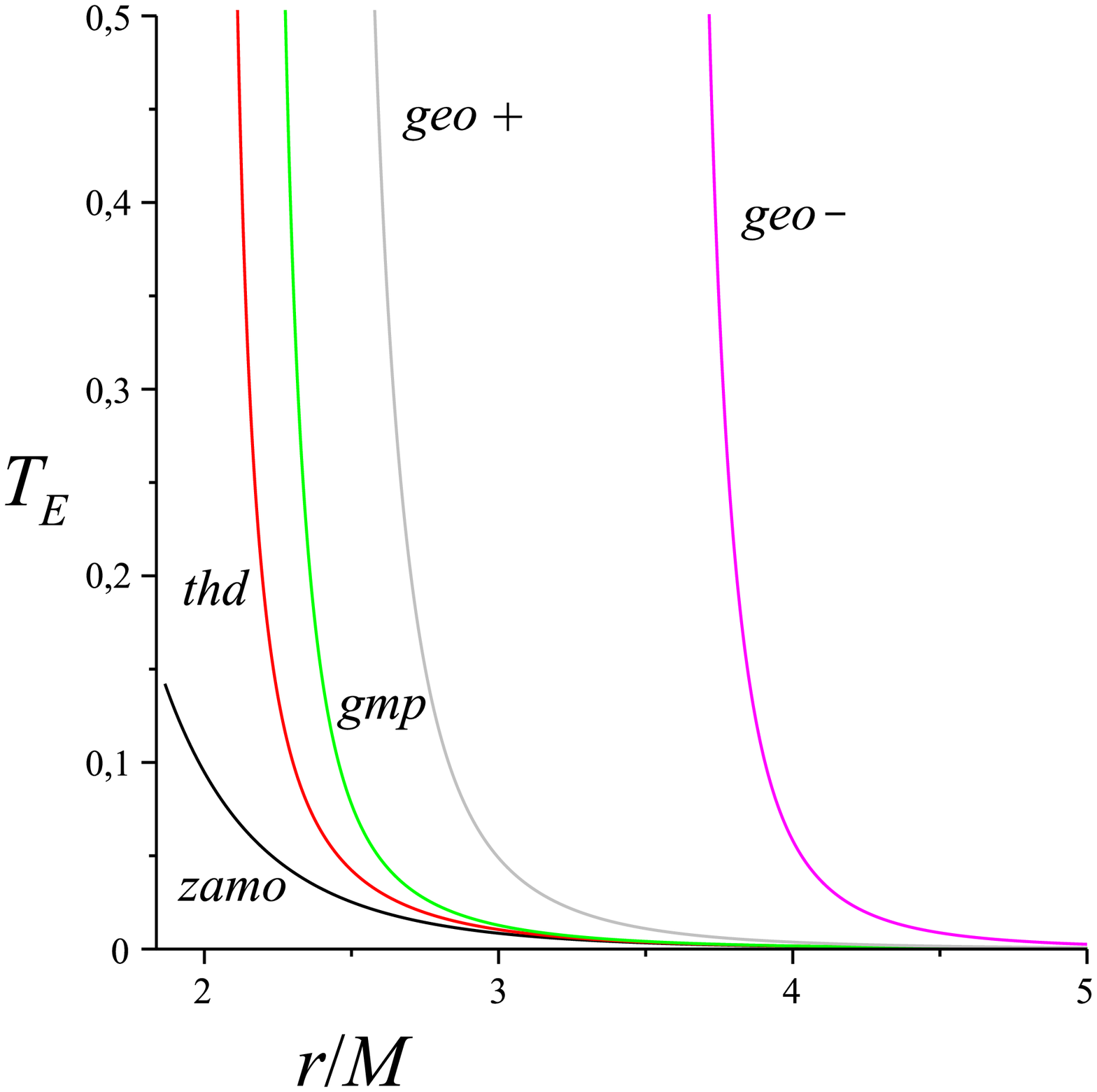}&\quad
\includegraphics[scale=0.28]{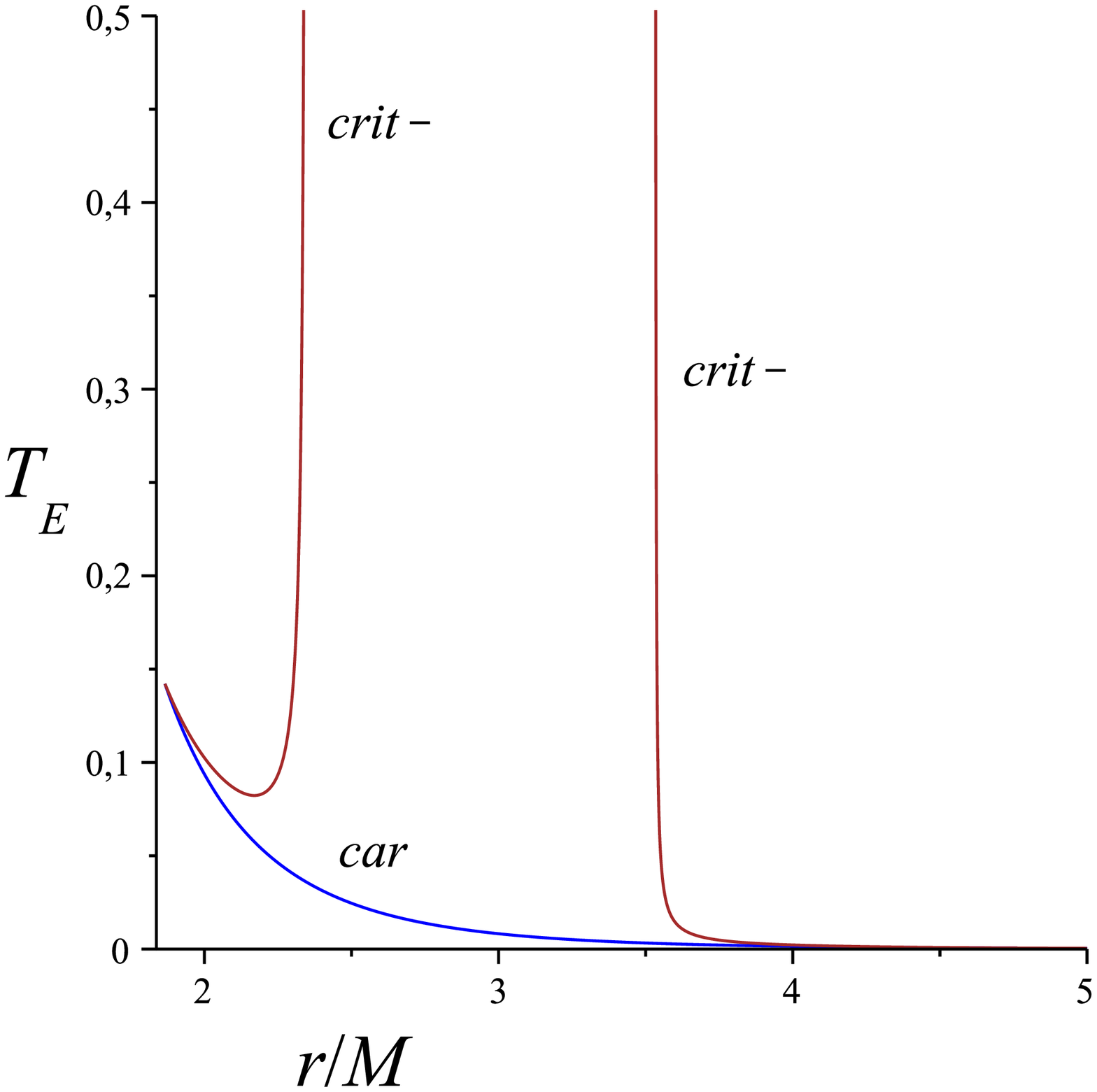}\\[.4cm]
\quad\mbox{(a)}\quad &\quad \mbox{(b)}\\[.6cm]
\includegraphics[scale=0.28]{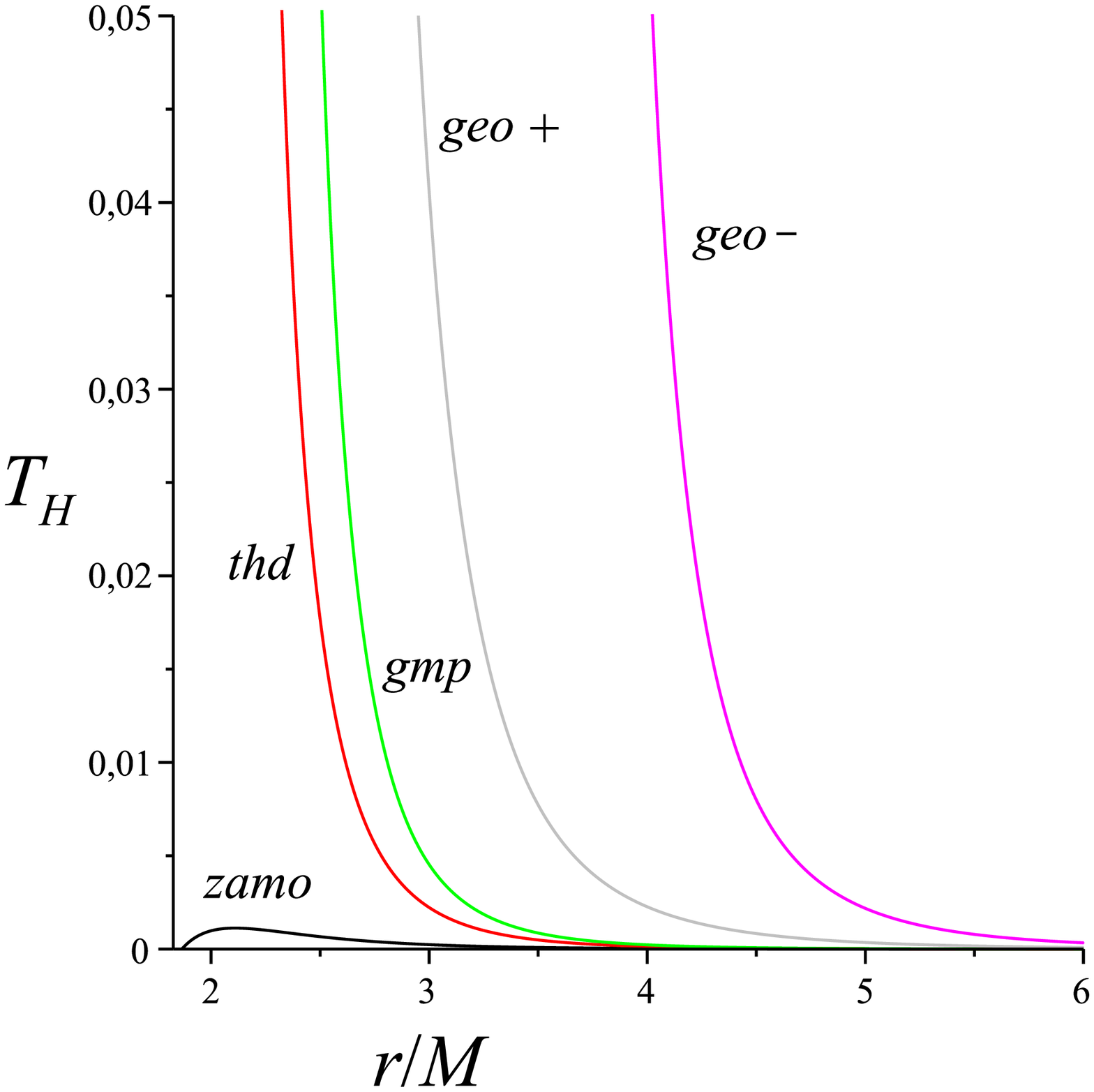}&\quad
\includegraphics[scale=0.28]{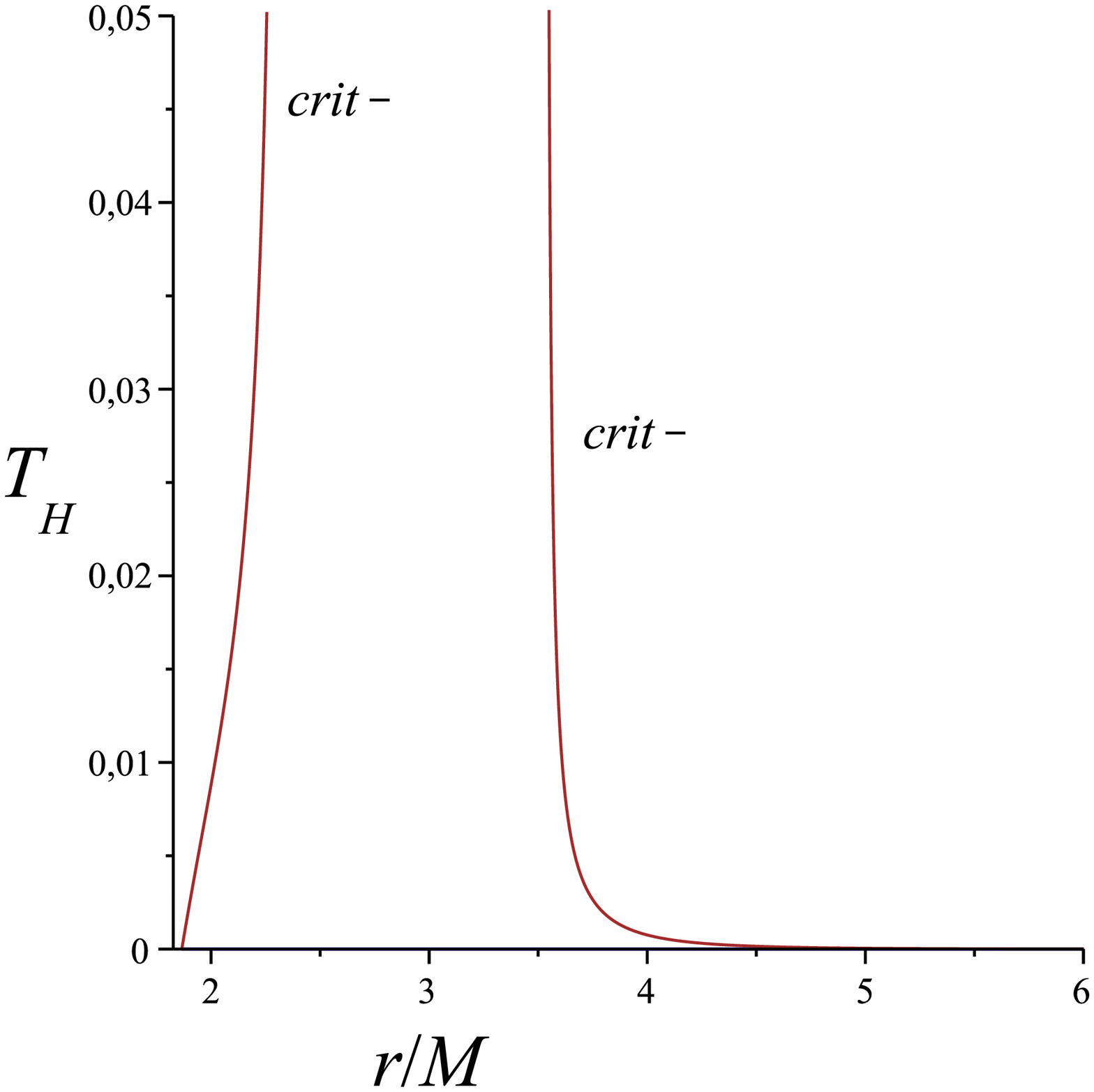}\\[.4cm]
\quad\mbox{(c)}\quad &\quad \mbox{(d)}
\end{array}$\\
\end{center}
\caption{The behaviors of the tidal indicators ${\mathcal T}_E(U)$ and ${\mathcal T}_H(U)$ as measured by different observers are shown as functions of the radial coordinate for the choice $a/M=0.5$.
Units on the vertical axis are chosen so that $M=1$.   
}
\label{fig:3}
\end{figure}

We list below the expression for ${\mathcal T}_H(U)$ corresponding to the different families of observers introduced in the previous section:
\begin{eqnarray}\fl\quad
{\mathcal T}_H(n)&=&\frac{18a^2M^2}{r^8}\frac{(r^2+a^2)^2\Delta}{(r^3+a^2r+2a^2M)^2}\,,\nonumber\\
\fl\quad
{\mathcal T}_H(u_{\rm(thd)})&=&\frac{18a^2M^2}{r^8}\frac{\Delta}{(r-2M)^2}\,,\nonumber\\
\fl\quad
{\mathcal T}_H(u_{\rm(car)})&=&0\,,\nonumber\\
\fl\quad
{\mathcal T}_H(u_{\rm(geo)\pm})&=&\frac{18M^2}{r^7}\frac{\Delta\left[ \sqrt{M}(r^2-3Mr+2a^2)\mp a\sqrt{r}(r-M) \right]^2}{(9M^2r-4a^2M-6Mr^2+r^3)^2}\,,\nonumber\\
\fl\quad
{\mathcal T}_H(u_{\rm(gmp)})&=&\frac{18a^2M^2}{r^4}\frac{\Delta(r+M)^2}{(r^4-2r^3M-2a^2Mr-M^2a^2)^2}\,,\nonumber\\
\fl\quad
{\mathcal T}_H(u_{\rm(crit)-})&=&\frac{18a^2M^2}{r^7}\frac{\Delta}{9M^2r-4a^2M-6Mr^2+r^3}\,.
\end{eqnarray}
All the magnetic-type tidal indicator vanish at the horizon (where $\Delta =0$) and in the limit $r\to \infty$ (i.e., far enough from the source).
Their behaviors as functions of the radial coordinate are shown in Fig. \ref{fig:3}.

In the weak field limit $M/r\ll1$ the above expressions have the following asymptotic forms (up to the order $(M/r)^{10}$)
\begin{eqnarray}\fl\quad
{\mathcal T}_H(n)&\simeq&\frac{18a^2M^2}{r^8}\left(1-\frac{2M}{r}\right)\,,\nonumber\\
\fl\quad
{\mathcal T}_H(u_{\rm(thd)})&\simeq&\frac{18a^2M^2}{r^8}\left(1+\frac{2M}{r}\right)\,,\nonumber\\
\fl\quad
{\mathcal T}_H(u_{\rm(geo)\pm})&\simeq&
\frac{18M^3}{r^7}\left\{1+\frac{M}{r}\left(4+\frac{a^2}{M^2}\right)+\frac{M^2}{r^2}\left(15+13\frac{a^2}{M^2}\right)\right.\nonumber\\
\fl\quad
&&\left.\mp\left[2\frac{a}{M}\sqrt{\frac{M}{r}}+12\frac{a}{M}\left(\frac{M}{r}\right)^{3/2}+2\frac{a}{M}\left(29+3\frac{a^2}{M^2}\right)\left(\frac{M}{r}\right)^{5/2}\right]\right\}\,,\nonumber\\
\fl\quad
{\mathcal T}_H(u_{\rm(gmp)})&\simeq&{\mathcal T}_H(u_{\rm(crit)-})
\simeq\frac{18a^2M^2}{r^8}\left(1+\frac{4M}{r}\right)\,.
\end{eqnarray}

% figure 2

% figure 3

\subsection{Observer carrying a spin and measuring curvature}

Consider now a spinning test particle moving along a circular orbit in the equatorial plane of a Kerr spacetime. 
According to the Mathisson-Papapetrou model \cite{math37,papa51} supplemented by Dixon conditions (see, e.g., Ref. \cite{clockkerr} and references therein) such a motion is compatible with an
angular velocity that is spin-dependent, the spin vector $S=-se_{\hat\theta}$ being constant and orthogonal to the motion plane.
The corresponding linear velocity is given by
\begin{eqnarray}
\label{nuspin}
\nu&=&\nu_{\rm(geo)\pm}+{\hat s}\tilde\nu\,, \nonumber\\
\tilde\nu&=&-\frac32\frac{M^2}{r\sqrt{\Delta}}\frac{(r^3+a^2r+2a^2M)(r^3+a^2M+2a^2r)}{(r^3-a^2M)^2}\nonumber\\
&&\times\left[1\mp\frac{a}{\sqrt{Mr}}\frac{r^3+a^2M+2Mr^2}{r^3+a^2M+2a^2r}\right]\,,
\end{eqnarray}
where ${\hat s}\equiv s/(mM)\ll1$ is the spin parameter such that the length scale $s/m$ associated with the spinning body be much smaller than the natural length scale $M$ of the background gravitational field in order to avoid backreaction effects.

\begin{figure} 
\typeout{*** EPS figure 4}
\begin{center}
\includegraphics[scale=0.32]{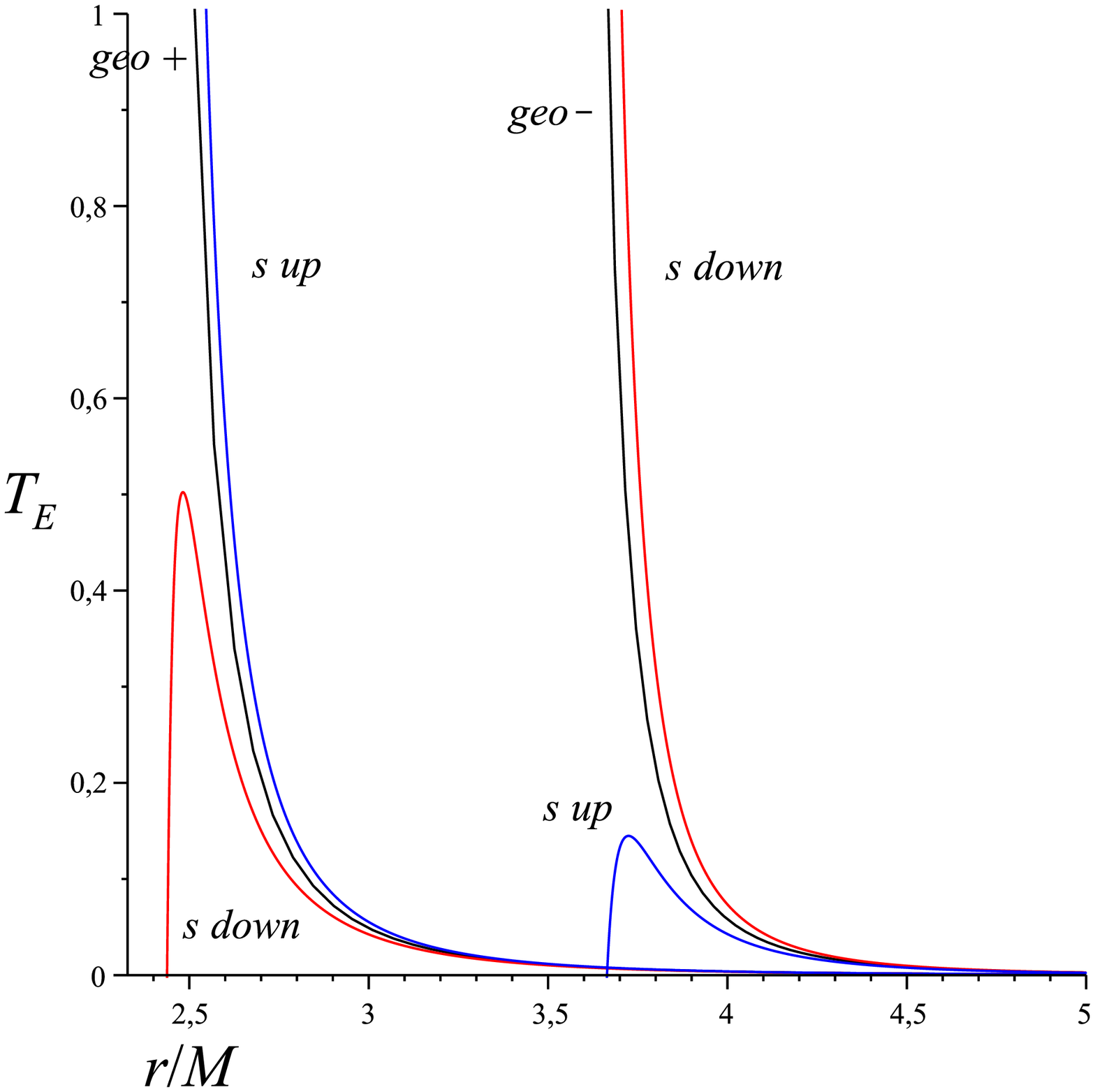}
\end{center}
\caption{The behavior ${\mathcal T}_E(U)$ is shown as a function of the radial coordinate in the case of a spinning particle with ${\hat s}=\pm0.1$ for the choice $a/M=0.5$.
Units on the vertical axis are chosen so that $M=1$.   
}
\label{fig:4}
\end{figure}

Substituting Eq. (\ref{nuspin}) in Eq. (\ref{tidalind2}) and retaining terms up to first order in $\hat s$ gives
\begin{eqnarray}\fl
{\mathcal T}_H(u)
&\simeq&{\mathcal T}_H(u_{\rm(geo)\pm})\left[1+2{\hat s}\tilde\nu\left(\frac{1}{1+\nu_\pm}+\frac{1}{1-\nu_\pm}+\frac{1}{\nu_\pm-\nu_{\rm(car)}}+\frac{1}{\nu_\pm-\bar\nu_{\rm(car)}}\right)\right]\nonumber\\
\fl
&=&{\mathcal T}_H(u_{\rm(geo)\pm})\left\{1+{\hat s}\left[
12a\frac{M^2}{r^2}\frac{\Delta}{9M^2r-4a^2M-6Mr^2+r^3}\right.\right.\nonumber\\
\fl
&&\left.\left.\mp3\left(\frac{M}{r}\right)^{3/2}\frac{(r-M)(r^2-3Mr+2a^2)}{9M^2r-4a^2M-6Mr^2+r^3}
\right]\right\}\,.
\end{eqnarray}
Fig. \ref{fig:4} shows the behavior of ${\mathcal T}_E(u)$ as a function of the radial coordinate.
If the structure of the body is taken into account ${\mathcal T}_E(u)$ can be made vanishing.

% figure 4

\section{Concluding remarks}

We have discussed the observer-dependent character of tidal effects associated with the electric and magnetic parts of the Riemann tensor with respect to an arbitrary family of observers in a generic spacetime.
Our considerations have then be specialized to the Kerr spacetime and to the family of stationary circularly 
rotating observers on the equatorial plane.
This family includes observers which have special properties: static, ZAMO, geodesic, geodesic meeting point, extremely accelerated, etc., and for each of them we have evaluated certain tidal indicators built up through the electric and magnetic parts of the Riemann tensor.
We have also considered the transformation laws of such tidal indicators when passing from one observer to another.
Our analysis has shown in particular that no equatorial plane circularly rotating observer in the Kerr spacetime can measure a vanishing electric tidal indicator, whereas it is the family of Carter's observers that measures a zero magnetic one.

\appendix

\section{Rapidity parametrization of tidal indicators} 

The expression for the tidal indicator ${\mathcal T}_E(U)$ given by Eq. (\ref{tidalind2}) can be cast in a more compact form if one introduces the rapidity parameter $\alpha$ instead of $\nu$, according to the relations
\beq
\nu=\tanh \alpha\,, \quad
\gamma=\cosh\alpha\,, \quad
\gamma \nu=\sinh \alpha\,.
\eeq
In fact, a quantity of the form 
\beq
{\mathcal F}(\nu,\nu_*)=\gamma^2\left(\nu^2+\frac{2\nu}{\nu_*}+1\right)
\eeq
where $\nu_*$ is a generic parameter, 
can be equivalently written as
\begin{eqnarray}
\label{eqA3}
{\mathcal F}(\alpha,\nu_*)&=&\sinh^2\alpha +\frac{2}{\nu_*} \sinh \alpha \cosh\alpha + \cosh^2\alpha\nonumber\\
&=& \cosh 2\alpha +\frac{1}{\nu_*} \sinh 2\alpha \,.
\end{eqnarray}
Next, denoting by
\beq
\frac{1}{\nu_*}=\tanh 2\beta_*\,,
\eeq
Eq. (\ref{eqA3}) becomes
\begin{eqnarray}
{\mathcal F}(\alpha,\nu_*)
&=& \cosh 2\alpha +\tanh 2\beta_* \sinh 2\alpha =\frac{\cosh 2(\alpha+\beta_*)}{\cosh 2\beta_*} \,.
\end{eqnarray}
Therefore, for instance, from  the first of Eqs. (\ref{tidalind2}) we have
\begin{eqnarray}
\label{tidalind3}
{\mathcal T}_E(U)
&=& {\mathcal T}_E(n)\, {\mathcal F}(\nu,\nu_{\rm(rel)-})\, {\mathcal F}(\nu,-\nu_{\rm(rel)+})\,\nonumber\\
&=& {\mathcal T}_E(n)\frac{\cosh [2(\alpha+\beta_{\rm(rel)-})]\cosh [2(\alpha-\beta_{\rm(rel)+})]}{\cosh [2\beta_{\rm(rel)-}]\cosh [2\beta_{\rm(rel)-}]} \nonumber\\
&=& {\mathcal T}_E(n)\frac{\cosh\left[4\left(\alpha-\sigma_+\right)\right] +\cosh \left(4\sigma_-\right)}{\cosh (4\sigma_-)+\cosh 4\sigma_+}\,,
\end{eqnarray}
where
\beq
\sigma_\pm = \frac{\beta_{\rm(rel)+}\pm \beta_{\rm(rel)-}}{2}\,.
\eeq
Introducing the rapidity parametrization in the expression (\ref{nureldef}) for $\nu_{\rm(rel)\pm}$ leads to
\beq
\frac{1}{\nu_{\rm(rel)\pm}}\equiv \tanh (2 \beta_{\rm(rel)\pm})
=\tanh 2(\tilde \alpha \mp \alpha_{\rm (car)})\,,
\eeq
where
\beq
\nu_{\rm (car)}=\tanh \alpha_{\rm (car)}\,,\qquad \frac12=\tanh 2\tilde  \alpha\,, 
\eeq
so that $\tilde \alpha=\frac14 \ln 3$ and
\beq
\tilde \alpha \mp \alpha_{\rm (car)} =  \beta_{\rm(rel)\pm }\,,
\eeq
whence
\beq
\sigma_+ =\tilde \alpha\,,\qquad \sigma_-=-\alpha_{\rm (car)}\,.
\eeq
Furthermore, it follows that 
\beq
\frac{\rmd {\mathcal T}_E(U)}{\rmd \alpha} \propto \sinh \left(4\left(\alpha-\sigma_+\right)\right)\,,
\eeq
and $\sigma_+$ is the value of $\alpha$ that extremizes ${\mathcal T}_E(U)$, corresponding to a physical observer or a timelike curve only in the case
$\tanh \sigma_+ <1$, a fact that in this case, with the corresponding form of $\sigma_+$, never happens.

\section*{Acknowledgments}
The authors thank ICRANet for support.


\section*{References}

\begin{thebibliography}{00}

\bibitem{abadieetal}
Abadie J {\it et al} 2010
{\it Class.\ Quantum Grav.} {\bf 27} 173001

\bibitem{newt1}
Lee W H and Kluzniak W 1999
{\it Astrophys. J.} {\bf 526} 178 

\bibitem{approxgr}
Faber J A,  Baumgarte T W, Shapiro S L, Taniguchi K and Rasio F A 2006
{\it Phys. Rev.} D {\bf 73} 024012 

\bibitem{fullgr1}
Shibata M and Ury$\rm\bar{u}$ K 2006
{\it Phys. Rev.} D {\bf 74} 121503(R); 
2007 {\it Class.\ Quantum Grav.} {\bf 24} S125

\bibitem{fullgr2}
Duez M D, Fourcart F, Kidder L E, Pfeiffer H P, Scheel M A and Teukolsky S A 2008
{\it Phys. Rev.} D {\bf 78} 104015

\bibitem{fullgr3}
Shibata M,  Kyutoku K,  Yamamoto T and Taniguchi K 2009
{\it Phys. Rev.} D {\bf 79} 044030

\bibitem{fla-hin}
Flanagan E E and Hinderer T 2008
% Constraining NS tidal Love numbers with gravitational wave detectors
{\it Phys. Rev.} D {\bf 77} 021502 

\bibitem{hin1}
Hinderer T 2008
% Tidal Love numbers of neutron stars
{\it Astrophys. J.} {\bf 677} 1216; 
erratum 2009 {\it Astrophys. J.} {\bf 697} 964

\bibitem{love}
Love A E H 1911
{\it Some Problems of Geodynamics} (Ithaca, USA: Cornell University Library)

\bibitem{Bin-poi}
Binnington T and Poisson E 2009
% Relativistic theory of tidal Love numbers
{\it Phys. Rev.} D {\bf 80} 084018 

\bibitem{damour1}
Damour T and Nagar A 2010
{\it Phys. Rev.} D {\bf 81} 084016 

\bibitem{mash75}
Mashhoon B 1975
% On tidal phenomena in a strong gravitational field
{\it Astrophys. J.} {\bf 197} 705; 
1977
%Tidal radiation
{\it Astrophys. J.} {\bf 216} 591.

\bibitem{dixon64} 
Dixon W G 1964  
{\it Il Nuovo Cimento} {\bf 34}  317 

\bibitem{dixon69}
Dixon W G 1970 
%``Dynamics of extended bodies in general relativity I. 
%  Momentum and angular momentum",
{\it Proc.\ R.\ Soc.} A {\bf 314}  499 

\bibitem{dixon70}
Dixon W G 1970 
%``Dynamics of extended bodies in general relativity II. 
%  Moments of the charge-current vector",
{\it Proc.\ R.\ Soc.} A {\bf 319}  509 

\bibitem{dixon73}
Dixon W G 1973 
{\it Gen.\ Rel.\ Grav.} {\bf 4}  199 

\bibitem{dixon74}
Dixon W G 1974  
{\it Phil.\ Trans.\ R.\ Soc.} A {\bf 277}  59 

\bibitem{fishbone}
Fishbone L G 1973
{\it Astrophys. J.} {\bf 185} 43 

\bibitem{hiscock}
Hiscock W A 1977
{\it Astrophys. J.} {\bf 216} 908

\bibitem{marck}
Marck J A 1983
{\it Proc.\ R.\ Soc.} A {\bf 385} 431 

\bibitem{carter}
Carter B and Luminet J P 1983
{\it Astron. Astrophys.} {\bf 121} 97; 
1985 
{\it Mon. Not. R. Astron. Soc.} {\bf 212} 23 

\bibitem{marck2}
Marck J A, Lioure A and Bonazzola S 1996
{\it Astron. Astrophys.} {\bf 306} 666

\bibitem{shibata}
Shibata M 1996
{\it Prog. Theor. Phys.} {\bf 96} 917 

\bibitem{frolov}
Diener P,  Frolov V P, Khokhlov A M, Novikov I D  and Pethick C J 1997
{\it Astrophys. J.} {\bf 479} 164 

\bibitem{mino}
Ishii M, Shibata M  and  Mino Y 2005
{\it Phys. Rev.} D {\bf 71}  044017 

\bibitem{math37} 
 Mathisson M 1937
 {\it Acta Phys.\ Pol.} {\bf 6} 167 

 \bibitem{papa51} 
 Papapetrou A 1951
 {\it Proc.\ Roy.\ Soc.} {\bf 209} 248

\bibitem{mashsingh}
Mashhoon B and  Singh D 2006
{\it Phys.\ Rev.} D {\bf 74}  124006 

\bibitem{spindev_schw}
Bini D, Geralico A and Jantzen R  T 2011
% Spin-geodesic deviations in the Schwarzschild spacetime 
{\it Gen. Rel. Grav.} {\bf 43} 959 

\bibitem{spindev_kerr}
Bini D and Geralico A 2011
%Spin-geodesic deviations in the Kerr spacetime
{\it Phys.\ Rev.} D {\bf 84} 104012

\bibitem{strains1} 
Bini D,  de Felice F and Geralico A 2006
%Strains in General Relativity 
{\it Class.\ Quantum Grav.}  {\bf 23} 7603 

\bibitem{pirani56}
Pirani F A E 1957
%Invariant Formulation of Gravitational Radiation Theory
{\it Phys. Rev.} {\bf 105}  1089

\bibitem{mcclune}
Mashhoon B and McClune J  C  1993
{\it Mon. Not. R. Astron. Soc.} {\bf 262} 881

\bibitem{chicmash1}
Chicone C and Mashhoon B  2005 
{\it Ann. Phys. (Leipzig)} {\bf 14} 290 

\bibitem{chicmash2}
Chicone C and Mashhoon B  2006
{\it Class.\ Quantum Grav.} {\bf 23} 4021

\bibitem{strains2} 
Bini D, de Felice F and Geralico A 2007
%Strains and axial outflows in the field of a rotating black hole 
{\it Phys. Rev.} D {\bf 76}  047502 

\bibitem{idcf1}
Bini D, Carini P and Jantzen R T  1997 
{\it Int.\ J.\ Mod.\ Phys.} D {\bf 6}  1 

\bibitem{idcf2}
Bini D, Carini P and Jantzen R T  1997 
{\it Int.\ J.\ Mod.\ Phys.} D {\bf 6}  143 

\bibitem{bjm}
Bini D, Jantzen R T and Mashhoon B 2001  
{\it Class. Quant. Grav.} {\bf 18}  653 

\bibitem{bjdf} 
Bini D, de Felice  F and Jantzen  R  T 1999
{\it Class.\ Quantum Grav.} {\bf 16}  2105 

\bibitem{bjelba}
Bini D and Jantzen R T 2003
Special Observers in the Kerr Spacetime 
{\it Advances in General Relativity and Cosmology (in Memory of A Lichnerowicz, Elba, 2002)} 
ed G Ferrarese (Naples: Bibliopolis) pp 287-296

\bibitem{book}
de Felice F and Bini D 2010
\textit{Classical Measurements in Curved Space-times} (Cambridge: Cambridge University Press)

\bibitem{maar}
Maartens R 2008
{\it Gen. Rel. Grav.} {\bf 40} 1203 

\bibitem{carter68}
Carter B  1968
%Global structure of the Kerr family of gravitational fields
{\it Phys. Rev.} {\bf 174} 1559 

\bibitem{iyer-vish}
Iyer B R and Vishveshwara C V 1993
%The Frenet-Serret description of gyroscopic precession
{\it Phys.\ Rev.} D {\bf 48} 5706 

\bibitem{bmj}
Bini D,  Merloni A and Jantzen R T 1999 
{\it Class. Quantum Grav.} {\bf 16} 1333

\bibitem{fdfacc}
de Felice F 1994 
{\it Class.\ Quantum Grav.} {\bf 11}  1283 

\bibitem{semerak}
Semer\'ak O 1996 
{\it Gen.\ Rel.\ Grav.} {\bf 28}  1151 

\bibitem{ht68}
Hartle J B and Thorne K S 1968
{\it Astrophys. J.}  {\bf 153}  807 

\bibitem{clockkerr}
Bini D, de Felice F and  Geralico A 2004
{\it Class.\ Quantum Grav.} {\bf 21} 5441 


\end{thebibliography}
\end{document}